# Spinless Bosons or Spin 1/2 Fermions in a 1D Harmonic Trap with Repulsive Delta Function Interparticle Interaction I - General Theory

Zhong-Qi Ma[1] and C. N. Yang[2,*]

[1] Institute of High Energy Physics, Chinese Academy of Sciences, Beijing 100049, China
[2] Tsinghua University, Beijing, China and Chinese University of Hong Kong, Hong Kong

*Abstract:* In these two papers, we solve the N body 1D harmonically trapped spinless Boson or spin 1/2 Fermions with repulsive δ function interaction in the limit $N \rightarrow \infty$.

*Introduction*

Stimulated by recent experimental and theoretical works [1,2,3] we give in this and a following paper the *ground state* energy and particle density distribution for $N$ spinless Bosons or spin 1/2 Fermions trapped in a harmonic trap with repulsive delta function interparticle interaction in the limit that $N \rightarrow \infty$. The Hamiltonian for the system is

$$H = H_1 + \sum_{i=1}^{N} V(x_i), \quad V(x_i) = x_i^2/2. \tag{I.1}$$

$$H_1 = \sum_{i=1}^{N} -\frac{1}{2}\frac{\partial^2}{\partial x_i^2} + g\sum_{i>j}^{N} \delta(x_i - x_j), \quad (g \geq 0). \tag{I.2}$$

We shall show explicitly later that as $N \rightarrow \infty$, at any fixed $x$,

$$\text{particle density} \quad \rho(x) \sim 0(\sqrt{N}), \tag{I.3}$$

in the limit that

$$g/\sqrt{N} = \text{a finite number.} \tag{I.4}$$

*Thomas-Fermi Method*

(I.3) shows that as $N \to \infty$, the density of particles at a fixed position $x$ approaches infinity. In other words most of the particles at $x$ have very very short wave lengths. I.e. the system at $x$ becomes a quantum gas (or liquid) satisfying thermodynamic laws. *Thus the Thomas-Fermi method becomes exact in the limit that $N \to \infty$.* I.e. the local density approximation is asymptotically exact in the limit $N \to \infty$.

We now turn to the thermodynamics within a macroscopic but small region $dx$ at position $x$. There are $\rho(x)dx$ particles within this region, forming the ground state for Hamiltonian $H_1$ with $N$ replaced by $\rho(x)dx$. The energy of this small thermodynamic system in length $dx$ has been explicitly given by Lieb and Liniger [4] for spinless Bosons in 1963 and by Yang [5] for spin 1/2 Fermions in 1967. We shall denote this energy by $e_1 dx$. The total energy for the ground state for $H$ is then

$$E = \int e_1 dx + \int \rho(x) V(x) dx, \tag{I.5}$$

where according to [4] and [5]

$$e_1 = \rho g^2 \zeta(\rho/g). \tag{I.6}$$

Function $\zeta$ for spinless Bosons can be calculated from the Fredholm equation of reference [4]. For spin 1/2 Fermions, there is the additional complexity of symmetry of the space wave function under the permutation group $S_N$ [5,3,6]. $\zeta$ would then depend on an *intensive* parameter $\xi$ which describes the symmetry of the space wave functions [7]. Recently You [8] has recalculated $\zeta$ for both Bosons and Fermions and fitted the resultant functions $\zeta$ with approximate algebraic expressions.

We now study the *intensive* properties of the thermodynamic system in $dx$. To avoid confusion we expand this interval *homogeneously* into length $\mathsf{L}$, and define *extensive* quantities for the expanded length:

$$\mathscr{E}_1 = \mathsf{L} e_1, \quad \mathscr{n} = \mathsf{L} \rho. \tag{I.7}$$

Then, $$\mathscr{E}_1 = \mathscr{n} g^2 \zeta(\beta), \quad \beta = \rho/g = \mathscr{n}/(\mathsf{L} g). \tag{I.8}$$

Thermodynamics demands that

$$d\mathcal{E}_1 = -p\,dL + \mu\,d\mathcal{N} \tag{I.9}$$

where $p$ and $\mu$ are *intensive* quantities: the *pressure* and the *chemical potential*. Thus

$$p = g\rho^2 \zeta'(\beta) \tag{I.10}$$

and

$$\mu = g^2 \zeta(\beta) + g\rho\zeta'(\beta). \tag{I.11}$$

*Equilibrium*

Different $dx$ regions have different densities $\rho$ and therefore different values of $e_1$. For the ground state, we vary $\rho(x)$ to obtain the minimum for the total energy $E$ (which was given in (I.5)), under the condition that $\int \rho\,dx$ is fixed. Thus

$$\delta e_1 + V(x)\delta\rho + \lambda\delta\rho = 0.$$

I.e.
$$\frac{de_1}{d\rho} + V(x) + \lambda = 0, \text{ or } g^2\zeta(\beta) + g\rho\zeta'(\beta) = -\lambda - V(x), \tag{I.12}$$

where $\lambda$ is Lagrange's multiplier. Using (I.11) this means

$$\mu + V(x) + \lambda = 0 \tag{I.13}$$

which is the thermodynamic condition for equilibrium, as expected.

For spin 1/2 Fermions in this variational procedure we keep the parameter $\xi$ fixed at the same value for all $x$. This needs justification which will be given in paper II.

*"Barometric" Pressure Balance*

It is clear from (I.10) and (I.11) that

$$\frac{dp}{d\rho} = \rho \frac{d\mu}{d\rho}. \tag{I.14}$$

Thus $\quad \dfrac{dp}{dx} = \dfrac{dp}{d\rho}\dfrac{d\rho}{dx} = \rho \dfrac{d\mu}{dx}.$

But (I.13) shows that $\dfrac{d\mu}{dx} = -\dfrac{dV}{dx} = \text{Force per particle} \equiv F(x).$

Thus $\quad \dfrac{dp}{dx} = \rho F(x).$

$$\tag{I.15}$$

Or $\quad p_A - p_B = (\rho\, dx)F(x)$

which is illustrated in Fig. 1 showing "barometric" pressure building up toward the trap center due to the "weight" of the $\rho\, dx$ trapped atoms.

*$\rho$ Dependence on x*

Equation (I.13) is a relationship between $\mu(\rho)$ and $V(x)$. Can we invert this relationship to solve for a *unique* $\rho$ at any given $x$? The answer is yes, because $\dfrac{d\mu}{d\rho} = \dfrac{dp}{d\rho}\Big/\rho$, and $\dfrac{dp}{d\rho}$ is always $> 0$ from general principles. Thus $\dfrac{d\mu}{d\rho} > 0$ and $\mu = \mu(\rho)$ is a monotonic function of $\rho$ with values between $\mu(0)$ and $\mu(\infty)$. Within this range, $\rho$ is a single valued function of $\mu$, hence of $x$, according to (I.13).

We shall present in paper II the explicit numerical values, calculated from the results of reference [8], for the asymptotic density distribution $\rho(x)$ and for the asymptotic total energy, as functions of $g/\sqrt{N}$, in the limit that $N \to \infty$.

One of us (Ma) is partly supported by the National Science Foundation of China under Grants No. 10675050.

___________

*Electronic address: cnyang@tsinghua.edu.cn

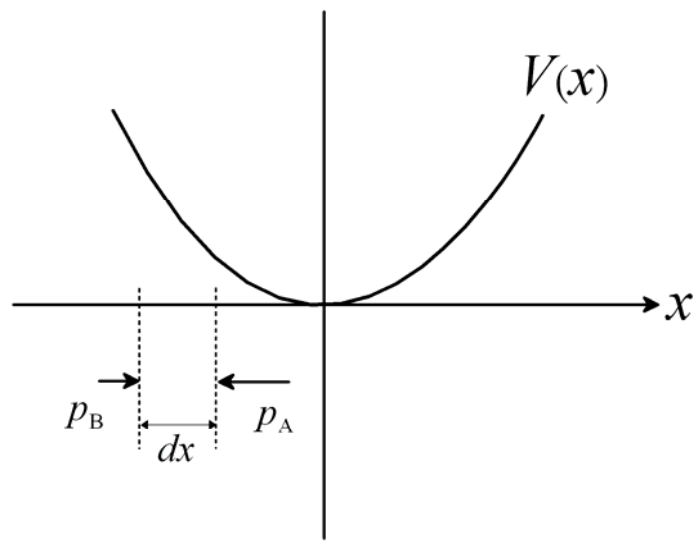

Figure 1.  Pressure Balance.  "Barometric" pressure $p_A$ is larger than $p_B$ to balance the "weight" of the $\rho dx$ particles in the interval $dx$.  See (I.15).